\documentclass
[twocolumn,showpacs,preprintnumbers,amsmath,amssymb,prc]{revtex4}
\usepackage{graphicx}
\usepackage{dcolumn}
\usepackage{bm}

\begin{document}

\title{Symmetry energy from nuclear multifragmentation}

\author{Swagata Mallik and Gargi Chaudhuri}
\affiliation{Theoretical Physics Division, Variable Energy Cyclotron Centre, 1/AF Bidhannagar,
Kolkata 700064, India}

\date{\today}

\begin{abstract}
The ratio of symmetry energy coefficient to temperature $C_{sym}/T$ is extracted from different
prescriptions using the isotopic as well as the isobaric yield distributions
 obtained in different projectile fragmentation reactions. It is found that the values
extracted from our theoretical calculation agree with those  extracted from the experimental data but they
differ very much from the input value of the symmetry energy used. The best possible way to deduce the value of
 the symmetry energy coefficient is to use the fragment yield at the breakup stage of the reaction and  it is better to use
the grand canonical model for the fragmentation analysis. This is because  the formulas that are used for the
deduction of the symmetry energy coefficient are all derived in the framework of the grand canonical ensemble which is valid
 only at the break-up (equilibrium) condition. The yield of "cold" fragments either from the theoretical models
 or from experiments when used for extraction of the symmetry energy coefficient using
these prescriptions might lead to the wrong conclusion.
\end{abstract}

\pacs{25.70Mn, 25.70Pq}
\maketitle

{\bf {\it Introduction:-}}
Isospin-dependent phenomena in nuclear physics has been an active area of research \cite{Bao-an-li1, Bao-an-li2} in recent years with the aim of enriching our knowledge about the symmetry term
 of the nuclear equation of state.  The study of this quantity at different regimes of density and temperature is a hot topic \cite{Tsang4, Lehaut, Samaddar} in the nuclear physics community.
 This term plays an important role in areas of astrophysical interest such as the study of supernova explosions and the properties of neutron stars \cite{Steiner}. This also has significant
influence in deciding the structure of neutron-rich and neutron-deficient nuclei. The study of nuclear multifragmentation in heavy-ion reactions is an important tool for extracting information
about the symmetry energy term, and this has created much interest in the nuclear physics community in recent years \cite{Souliotis, Shetty2, Fevre, Henzlova1, Shetty3, Shetty1, Chen, Huang, Marini, Ma1, Ma2}.\\
In nuclear multifragmentation reactions, the neutron-proton composition of the break-up fragments is dictated by the asymmetry term of the equation of state and hence the study of
 the multifragmentation process allows one to obtain information about the symmetry term. Statistical models \cite{Das1, Bondorf1, Gross1} that are simple and economic are very
 successful in predicting the outcome of nuclear multifragmentation reactions. Different formulas \cite{Huang, Tsang1, Ono, Raduta2} have been proposed in the literature that connect
 the measurable fragment isotopic and isobaric observables of multifragmentation reactions to the symmetry energy of excited nuclei and these have been applied to the analysis of heavy-ion collision data.
These formulas have all been  deduced using the grand canonical version of the nuclear multifragmentation model assuming an equilibrium scenario for the break-up stage of the
disintegrating system. They have been used to analyze experimental data from different projectile fragmentation as well as central collision
reactions and the extracted values for the symmetry energy coefficient  $C_{sym}$ ranges between 15 and 30 MeV \cite{Souliotis, Shetty2, Fevre, Henzlova1, Shetty3}.
It has been observed from model calculations that the fragment distributions of the  secondary fragments after evaporation vary from those at the primary (breakup)
stage and hence extraction of $C_{sym}$ values from experimental yields that correspond to the 'cold' secondary fragments might lead to the wrong conclusion.
 This is primarily because the formulas that are  used to extract the values of $C_{sym}$ from the experimental yields are all derived using the prescription
of the grand canonical ensemble, which is valid only at the break-up (primary stage) of the multifragmenting system.
  The yield of the fragments at the breakup stage of the reaction (before de-excitation by secondary decay) should only be used to extract information
 about $C_{sym}$  using the existing prescriptions. This is the main message of this paper and it has been established using projectile fragmentation reactions.
It has also been pointed out that theoretical simulations done by the grand canonical model are only expected to reproduce the value of  $C_{sym}$ used as input to the model.
 Calculations by canonical or any other ensemble that are otherwise better suited for describing intermediate energy nuclear reactions might lead to values widely different
 from the input value of $C_{sym}$ used.  Results from canonical \cite{Das1} and grand canonical ensemble \cite{Chaudhuri2} differ in general for finite nuclei \cite{Mallik4}.
 This work also presents a comparative analysis of the predictive power of the different existing formulas both at the primary stage and also after evaporation and their relative agreement with experimental data.

{\bf {\it Model description:-}}
For the study of the nuclear symmetry energy coefficient from projectile fragmentation reactions we use a model \cite{Mallik2, Mallik3, Mallik101} that consists of three stages: (i) abrasion, (ii) multifragmentation, and (iii) evaporation. In heavy-ion collisions,
if the beam energy is high enough then at a particular impact parameter three different regions are assumed to be formed: (i) the participant, (ii) the projectile like fragment (PLF) or projectile spectator, and (iii) the target like fragment (TLF) or target spectator. Here we are interested in  the fragmentation of the PLF.

In the abrasion part (first stage), we have calculated the PLF volume $V_s(b)$ at different impact parameters ($b$) using straightline geometry. By knowing $V_s(b)$, we can determine the average number of neutrons and protons present in the projectile spectator \cite{Mallik2} by assuming the same neutron-to-proton ratio of projectile spectator and projectile. The impact parameter dependence of temperature is considered as $T(b)=7.5-4.5[(A_s(b)/A_0)]$ where $A_s(b)$ and $A_0$ are mass numbers of projectile spectator and original projectile respectively. This temperature profile successfully reproduces \cite{Mallik3, Mallik101} the differential charge distribution, total charge and mass distribution, and isotopic distribution obtained in different experiments \cite{Ogul, Mocko1, Mocko2, Henzlova2}.

In the statistical models \cite{Das1, Bondorf1, Gross1} of nuclear multifragmentation, it is assumed that the hot nuclear system expands and then, due to density fluctuations, it breaks up (multifragments) into composites of different sizes. Thermodynamic equilibrium is assumed to be reached at this expanded freeze-out configuration, where the interaction between the composites become unimportant (except for the long-range Coulomb force). The disintegration of hot nuclear system into various composites can be calculated using different statistical ensembles. In our model, in general canonical ensemble is used as it is more appropriate than the grand canonical for finite nuclei calculations. Results from these ensembles differ in general though they are found to converge under certain conditions \cite{Mallik4} for finite nuclei. The multifragmentation calculation of each PLF (second stage) formed at different impact parameters is done separately by using the canonical thermodynamical model (CTM) \cite{Das1} which is based on equilibrium statistical mechanics and involves the calculation of partition functions. In this model the average number of composites with $N$ neutrons and $Z$ protons formed from a PLF ${N_{s},Z_{s}}$ is given by
\begin{eqnarray}
\langle n_{N,Z}\rangle_{c} & = & \omega_{N,Z}\frac{Q_{N_{s}-N,Z_{s}-Z}}{Q_{N_{s},Z_{s}}}
\end{eqnarray}
where $Q_{N_{s},Z_{s}}$ is the total partition function, which can be easily calculated using the recursion relations \cite{Chase}.

The grand canonical ensemble can also be used instead of the canonical, and the average number of composites in this ensemble is given by \cite{Chaudhuri2},
\begin{eqnarray}
\langle n_{N,Z}\rangle_{gc} & = & e^{\beta\mu_{n}N+\beta\mu_{p}Z}\omega_{N,Z}
\end{eqnarray}

The neutron and proton chemical potentials $\mu_{n}$ and $\mu_{p}$ are determined by the baryon and charge conservation conditions which amounts to solving the equations $N_{s}=\sum Ne^{\beta\mu_{n}N+\beta\mu_{p}Z}\omega_{N,Z}$ and $Z_{s}=\sum Ze^{\beta\mu_{n}N+\beta\mu_{p}Z}\omega_{N,Z}$.

In both the models, the partition function $\omega_{N,Z}$ of a composite having $N$ neutrons and $Z$ protons is a product of two parts and is given by
\begin{eqnarray}
\omega_{N,Z}=\frac{V}{h^{3}}(2\pi mT)^{3/2}A^{3/2}\times z_{N,Z}(int)
\end{eqnarray}

The first part is due to the translational motion and the second part $z_{N,Z}(int)$ is the intrinsic partition function of the composite. $V$ is the volume available for translational motion, which is the difference between the freeze-out volume and normal nuclear volume. For projectile fragmentation calculation the freeze-out volume is assumed to be three times the normal nuclear volume.

We now list the properties of the composites used in this work.  The proton and the neutron are fundamental building blocks, thus $z_{1,0}(int)=z_{0,1}(int)=2$, where 2 takes care of the spin degeneracy.  For deuteron, triton, $^3$He and $^4$He we use $z_{N,Z}(int)=(2s_{N,Z}+1)\exp[-\beta E_{N,Z}(gr)]$ where $\beta=1/T, E_{N,Z}(gr)$ is the ground-state energy of the composite and $(2s_{N,Z}+1)$ is the experimental spin degeneracy of the ground state.  Excited states for these very-low-mass nuclei are not included.
For mass number $A\ge5$ we use the liquid-drop formula.  For nuclei in isolation, this reads\\
\begin{eqnarray}
z_{N,Z}(int)
&=&\exp\frac{1}{T}[W_0A-\sigma(T)A^{2/3}-a^{*}_c\frac{Z^2}{A^{1/3}}\nonumber\\
&&-C_{sym}\frac{(N-Z)^2}{A}+\frac{T^2A}{\epsilon_0}]
\end{eqnarray}

The expression includes the volume energy [$W_0=15.8$ MeV], the temperature dependent surface energy
[$\sigma(T)=\sigma_{0}\{(T_{c}^2-T^2)/(T_{c}^2+T^2)\}^{5/4}$ with $\sigma_{0}=18.0$ MeV and $T_{c}=18.0$ MeV], the Coulomb energy [$a^{*}_c=0.31a_{c}$ with $a_{c}=0.72$ MeV and Wigner-Seitz correction factor 0.31 \cite{Bondorf1}] and the symmetry energy ($C_{sym}=23.5$ MeV).  The term $\frac{T^2A}{\epsilon_0}$ ($\epsilon_{0}=16.0$ MeV) represents contribution from excited states since the composites are at a non-zero temperature. Different equations have been deduced using the grand canonical model which relates this symmetry energy coefficient $C_{sym}$ to temperature as well as the source and fragment compositions. This is dealt with in the next section.

The excited fragments produced after multifragmentation decay to their stable ground states. They can $\gamma$-decay to shed energy but may also decay by light particle emission to lower mass nuclei.  We include emissions of $n,p,d,t,^3$He and $^4$He. Particle-decay widths are obtained using Weisskopf's evaporation theory. Fission is also included as a de-excitation channel though for the nuclei of $A{\textless}100$ its role will be quite insignificant. The details of the evaporation stage (last stage) are described in Ref. \cite{Mallik1}.\\

{\bf {\it Symmetry energy from different formulas:-}}
There exist different formulas in the literature from which the symmetry energy coefficient has been extracted. Here we make a short review of these existing formulas.

The formula that connects the symmetry energy coefficient with the isospin asymmetry of the source was first proposed in the framework of the Expanding evaporating source (EES) model \cite{Tsang1}. This is based on the grand canonical ensemble, and assuming thermodynamic equilibrium at the time of fragmentation of two systems having same charge $Z_0$ but different masses $A_{01}$, $A_{02}$ ($A_{01}<A_{02}$) at the same temperature T, $C_{sym}$ is given by
\begin{equation}
C_{sym}(Z)=\frac{\alpha(Z) T}{4 \left[ \left( \frac {Z_0}{A_{01}}\right)^2-
\left( \frac {Z_0}{A_{02}}\right)^2\right] }.
\label{eq:csym_tsang}
\end{equation}
where $\alpha(Z)$  is the isoscaling parameter \cite{Tsang1, Botvina1, Chaudhuri4} of fragments having charge $Z$ which can be measured from the ratio of the isotopic yields \cite{Tsang2}. The suffix 0 indicates the fragmenting source. This formula, which is referred to as the isoscaling (source) formula, has been extensively used on experimental data \cite{Souliotis, Shetty2, Fevre, Henzlova1, Shetty3}.

The isoscaling (fragment) formula also derived \cite{Ono} from the grand canonical ensemble assumes equilibrium at the breakup stage of two fragmenting sources in identical thermodynamical states that differ in their isospin content (different masses $A_{01}$, $A_{02}$ but same charge $Z_0$). This is given by
\begin{equation}
C_{sym}(Z)=\frac{\alpha(Z) T}{4 \left[ \left( \frac {Z}{<A_1>}\right)^2-
\left( \frac {Z}{<A_2>}\right)^2\right] }
\label{eq:csym_ono}
\end{equation}

In this expression it is assumed that the isotopic distributions are essentially Gaussian and that the free energies contain only bulk terms. Here, $\alpha(Z)$ is the isoscaling parameter of fragments having $Z$ protons, $<A_{i}>$ is the average mass number of a fragment of charge $Z$ produced by source $i$=1 (less neutron rich), 2 (more neutron rich), and $T$ is the temperature of the decaying systems. This formula, which is  similar in structure to the isoscaling (source) formula, connects the symmetry energy coefficient to the isotopic composition of fragments instead of the isotopic composition of sources as in Eq. (5).

An alternate expression (fluctuation formula) has been derived in Ref. \cite{Raduta2, Chaudhuri3} connecting
the symmetry energy of a cluster of size $A$ to the width of its isotopic distribution.
Indeed, a Gaussian approximation on the grandcanonical expression for cluster yields
gives
\begin{equation}
C_{sym}(A)=\frac{AT}{2\sigma_I^2(A)},
\label{eq:csym_fluct}
\end{equation}
where $\sigma_I^2(A)$ indicates the width of the isobaric distribution of a cluster of size $A$ and $I=A-2Z$.

According to the isobaric yield ratio method \cite{Huang, Marini, Ma1, Ma2, Tsang3}, obtained from the grand canonical expression for cluster yields, $C_{sym}$ can be expressed as
\begin{equation}
C_{sym}(A)=-\frac{AT}{8}[ln{R(3,1,A)}-ln{R(1,-1,A)}]
\end{equation}
where, $R(I+2,I,A)$ is the yield ratio between two isobars differing by 2 units in isospin content as $R(I+2,I,A)=Y(I+2,A)/Y(I,A)$. This equation assumes the Coulomb terms in $ln{R(3,1,A)}$ and $ln{R(1,-1,A)}$ are same and in both ratios mixing entropy terms are neglected \cite{Tsang3}.\\

{\bf {\it Results:-}}
\begin{figure}[b]
\includegraphics[width=3.0in,height=3.0in,clip]{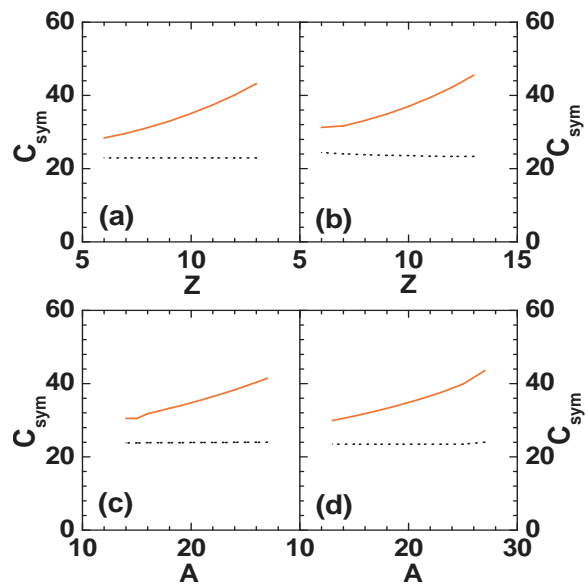}
\caption{ (Color online) Symmetry energy coefficient obtained from canonical (red solid lines) and grand canonical model (black dotted lines) at $T=$5 MeV and input $C_{sym}=$23.5 MeV. (a) and (b) represents the variation of extracted $C_{sym}$ with proton number $Z$ by using Eq. (5)  and Eq. (6)
 respectively from two sources having same $Z_0=25$ but different $A_0=60$ and $55$ where as (c) and (d) indicates the variation of extracted $C_{sym}$ with
mass number $A$ calculated by using Eqs. (7) and (8) for $Z_0=25$ and $A_0=55$. }
\label{fig1}
\end{figure}
The isotopic and isobaric yield distributions are related to the nuclear symmetry energy coefficient through
different prescriptions, which have already been described briefly in the previous section. These relations are derived using the yields of the fragments  obtained in the grand canonical ensemble. We first compare  the symmetry energy coefficient obtained using the different prescriptions in both the canonical and the grand canonical ensemble. In this calculation we have used the statistical models (canonical or grand canonical) in order to obtain the yields of the composites formed after multifragmentation (second stage) of the hot single source. The sources used for the first two methods [Fig 1(a) and 1(b)] are $A_{01}=55$ and $A_{02}=60$ and $Z_0=25$, while for the later two methods [Fig 1(c) and 1(d)]
 where a single source is required the source $A_0 =55$ and $Z_0=25$ is used. The temperature used for the calculation is $5$ MeV. These are the results for the primary fragments and no secondary decay is
used for the de-excitation of the excited primary fragments. There are some differences in the results from the canonical and the grand
 canonical ensemble, and these differences are almost same in all the methods used. In the canonical ensemble, the extracted symmetry energy coefficient changes
 with the fragment mass or charge and this variation is more or less same for all the four methods used for extraction of this coefficient. The value of extracted symmetry energy coefficient varies between 25 and 50 MeV while the input symmetry energy coefficient used is fixed at 23.5 MeV.

In contrast, in the results from the grand canonical ensemble, $C_{sym}$ is independent of the fragment mass or charge. The value of $C_{sym}$ extracted
from the grand canonical ensemble lies between $23$ and $24$ MeV and hence matches almost exactly with the input value used for the calculation.
This difference in results between two ensembles is mainly because these formulas are all derived using the prescription of the grand canonical ensemble and hence when this ensemble is
used to extract the value, the results agree with the input value. The results from the canonical ensemble deviate from that of the grand canonical ensemble in general for finite nuclei. Hence extraction of $C_{sym}$ leads to values that can differ widely from the input value used.
\begin{figure}[h]
\includegraphics[width=3.0in,height=1.8in,clip]{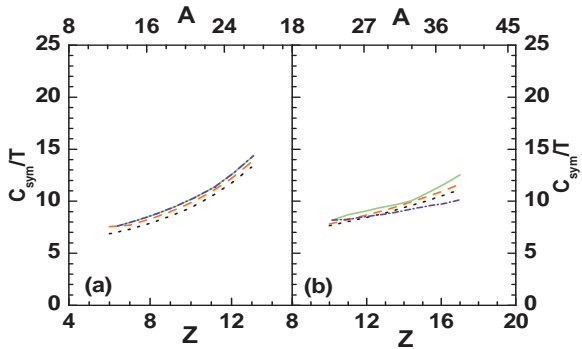}
\caption{ (Color online) Comparison of $C_{sym}/T$ calculated by using (i) Eq. (5) (black dotted lines) (ii) Eq. (6) (red dashed lines) (iii) Eq. (7) (green solid lines) and (iv) Eq. (8) (blue dash dotted lines) for primary fragments in projectile fragmentation reactions $Ni$ on $Be$ (left panel) and $Xe$ on $Pb$ (right panel). For (i) and (ii), $C_{sym}/T$ is calculated from two sets of reactions $^{58}$Ni on $^{9}$Be and $^{64}$Ni on $^{9}$Be (left panel) and $^{124}$Xe on $^{208}$Pb and $^{136}$Xe on $^{208}$Pb (right panel) and plotted against $Z$ where as for (iii) and (iv), $C_{sym}/T$ is calculated for only neutron less reactions and plotted against $A$. }
\label{fig2}
\end{figure}

In the next part, we show the results of the calculations from the projectile fragmentation model for two different reactions. Since for finite nuclei, the canonical ensemble is physically more acceptable than the grand canonical ensemble, in our model (as  described previously) we have used the canonical ensemble for the fragmentation
of the excited PLF. In this case we have used the reduced symmetry energy coefficient which is nothing but the ratio of the symmetry energy coefficient to temperature i.e, $C_{sym}/T$. Recent studies on this parameter using different methods can be found in ref. \cite{Chen, Huang, Marini, Ma2}. This is mainly because it is difficult to estimate temperature both from experimental yields and from the model results. Hence it is better to express the results in the form of $C_{sym}/T$ instead of using only $C_{sym}$ since unambiguous extraction of temperature is very difficult.  In the isoscaling (source) formula [Eq. (5)], symmetry energy is related to the isoscaling parameter $\alpha$  and the $Z/A$ value of the two sources. The other three formula depend solely on the properties of the fragments. In the isoscaling (fragment) formula [Eq. (6)], $C_{sym}/T$ depends on the isoscaling parameter and on the $Z/{\textless}A{\textgreater}$ values of the fragments. In the fluctuation formula [Eq. (7)], the coefficient depends on the width of the isobaric distribution of the fragments and their mass  while in the isobaric yield ratio [Eq. (8)], it depends on the ratio of the fragment yields and their mass. In the break-up stage of the multifragmentation reaction, the yields of the primary fragments can be used to deduce the values of $C_{sym}/T$ from all the four formulas.  The projectile fragmentation reactions involved are $^{58}$Ni on $^{9}$Be and $^{64}$Ni on $^{9}$Be at 140 MeV/n \cite{Mocko1, Mocko2}. From Eqs. (5) and (6), the variation of the reduced symmetry energy coefficient $C_{sym}/T$ with fragment charge $Z$ is shown in Fig. 2(a). In  the same figure the results from Eq. (7) and (8) are shown as functions of the fragment mass $A$. The first two methods [Eq. (5) and (6)] depend on the isoscaling parameter which is calculated using two  projectiles $Ni^{58}$ and $Ni^{64}$ having same value of Z. The fluctuation method [Eq. (7)] and the isobaric yield ratio method [Eq. (8)] depend on one source for the calculation of the reduced symmetry energy coefficient and the projectile $Ni^{58}$ has been used. It is seen that the results from the primary fragments are close to each other for all four formula and $C_{sym}/T$ increases with the fragment mass or charge. Since these values are for the break-up stage before the final de-excitation, no comparison has been made to the experimental data.

We have repeated the same calculation [Fig. 2(b)] for different projectiles ($Xe^{124}$ and $Xe^{136}$). The projectile fragmentation reactions involved were $^{124}$Xe on $^{208}$Pb and $^{136}$Xe on $^{208}$Pb at 1 GeV/n \cite{Henzlova2}. The beam energy is 1 GeV/n which is much higher than the previous value of 140 MeV/n \cite{Mocko1, Mocko2}. The trend of the results remains almost the same irrespective of the beam energy. The results from all four formulas are close to each other for the primary fragments (at the break-up stage) for a wide range of beam energies from 140 MeV/n to 1 GeV/n.

In the next stage we investigated the results for the secondary fragments i.e, after the evaporation of the excited primary fragments. In each figure, the results from the primary fragments are also shown for the sake of convenience in comparison. In Fig. 3(a) we have shown the result for the isoscaling (source) formula [Eq. (5)]. Here $C_{sym}/T$ decreases after evaporation because the isoscaling parameter $\alpha$ decreases after evaporation at the temperature range used here while the denominator of the right-hand side of Eq. (5), which depends only on the source sizes remain unchanged.  In  both results from the  model and from  the experimental data, the reduced symmetry energy coefficient does not depend very much on the fragment size. The  numbers obtained from the experimental data are less than those obtained from our model for the projectile fragmentation. In Fig. 3(b) we have plotted the results for the isoscaling (fragment) formula [Eq. (6)]. In this case, $C_{sym}/T$ increases after secondary decay and the results more or less agree with those extracted from the experimental data.
\begin{figure}[t]
\includegraphics[width=3.0in,height=3.0in,clip]{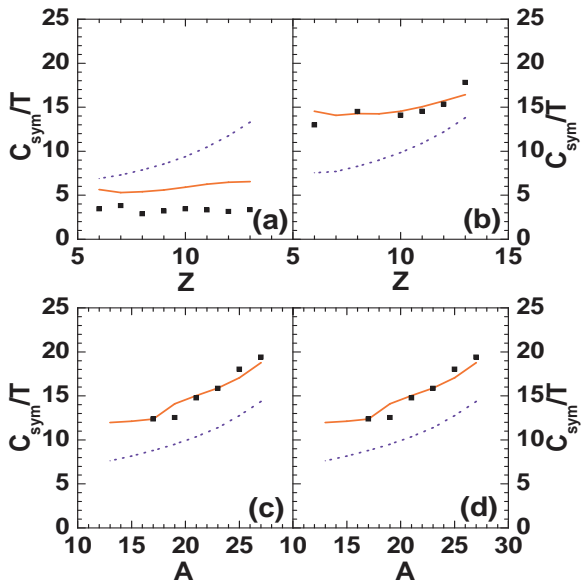}
\caption{ (Color online) Variation of $C_{sym}/T$ with atomic number $Z$ calculated by using (a) Eq. (5) and (b) Eq. (6) for $^{58}$Ni on $^{9}$Be and $^{64}$Ni
on $^{9}$Be reactions. Panels (c) and (d) depict the variation of $C_{sym}/T$ with mass number $A$ for $^{58}$Ni on $^{9}$Be reaction calculated from Eqs. (7) and
 (8) respectively. Experimental data (black squares) compared with theoretical results: primary  fragments (blue dotted lines) and secondary fragments (red solid lines). }
\label{fig3}
\end{figure}
\begin{figure}[b]
\includegraphics[width=3.0in,height=3.0in,clip]{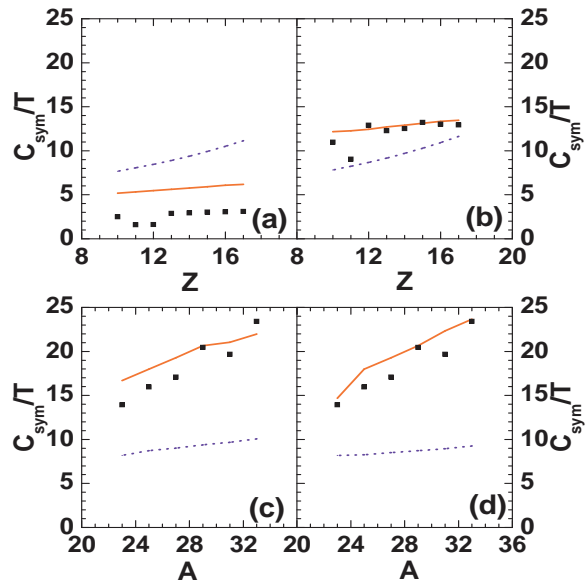}
\caption{ (Color online) Same as Fig.-3 except that here the projectile fragmentation reactions involved are $^{124}$Xe and $^{136}$Xe on $^{208}$Pb instead of $^{58}$Ni and $^{64}$Ni on $^{9}$Be. }
\label{fig4}
\end{figure}
The isoscaling (fragment) formula depends on the $Z/{\textless}A{\textgreater}$ values of the fragments [see Eq. (6)]. For the less
 neutron rich source,
 this quantity does not change very much after evaporation but for the more
neutron-rich nuclei, this quantity increases after evaporation since ${\textless}A{\textgreater}$ decreases as the  peak of the isotopic distribution shifts to the
 left (lower values of A) after evaporation. Hence the denominator of the right-hand side of
Eq. (6) decreases. The isoscaling parameter $\alpha$ in
 the numerator also decreases after evaporation but the denominator decreases much more and
 hence $C_{sym}/T$  increases after secondary decay as is seen from the Fig. 3(b). This is in contrast to the results from the isoscaling (source) formula where the denominator
 is independent of the property of the fragments.
In Fig. 3(c) we have plotted the results from the fluctuation formula [Eq. (7)]. Here also the trend of the results is same as in Fig. 3(b).  $C_{sym}/T$
 increases after secondary decay, and this is due to the fact that
$\sigma^2$ which is a measure of the width of the isobaric distribution, decreases after secondary decay and one can see from  Eq. (7) that
 if this decreases then  $C_{sym}/T$ will increase and this is exactly
what happens as we see from Fig. 3(c). The experimental values obtained in this case also are quite close to those obtained from the model.
In Fig. 3(d) we have plotted the results from the isobaric yield ratio method. In this case also the result is similar to that of the
 previous two cases and the reduced symmetry energy coefficient increases after
evaporation.  The numbers extracted from  the experimental data are close to those from the theoretical calculation. A similar trend of results
from the isobaric yield ratio method in Antisymmetrized molecular dynamics (AMD) model+GEMINI calculations is reported in \cite{Huang}.  In this case as seen
from Eq. (8), $C_{sym}/T$ depends on the ratios of yields of different isobars.
The reduced symmetry energy coefficient increases after evaporation because the width of the isobaric distribution decreases due to
secondary decay, which results in decrease of the yield $Y(-1,A)$
and  $Y(3,A)$ in Eq. (8) while the yield $Y(1,A)$ remains almost unchanged. In the results from Eq. (6)-(8),   $C_{sym}/T$ increases after evaporation,
 as compared to the results obtained from the primary fragments.

We have repeated this calculation for another projectile fragmentation reaction (Xe on Pb) at 1 GeV/n \cite{Henzlova2} and this is shown in Fig. 4.
The  observations and results are similar to those of the previous reaction in spite of the vast difference in the projectile energy and widely different target-projectile combination.

All the four formulas are derived from the yields of the grand canonical ensemble and they hold good for the breakup stage of the reaction.
 Hence any attempt to deduce the value of the symmetry energy coefficient
from the yields obtained after evaporation might lead to the wrong conclusion. Neither the experimental yields (which are the values after evaporation
 from the excited fragments) nor the yields of the secondary fragments
 from the model should be used to deduce the values of
the symmetry energy coefficient. It might be possible to deduce the value of $C_{sym}$ from
the break-up stage of the reaction, i.e., from the hot primary fragments, but
since it is difficult to access this stage from an experimental point of view, no attempts are made to do such calculations.\\

{\bf {\it Summary and conclusion:-}}
In this work we have compared the values of $C_{sym}$ obtained from  the primary fragments for the canonical and the grand canonical ensembles for a single source
 at a fixed temperature using the four different prescriptions.   The results from the canonical ensemble calculations show that the values obtained from the four formulas  agree with each other.
The results also display that for the canonical model, $C_{sym}$ varies with the fragment size and differs from the input $C_{sym}$ value
which is equal to 23.5 MeV. In the grand canonical model, this value is independent of the fragment size and is almost equal to the input value used.
 This is because  all the four formulas are deduced using the grand canonical ensemble at the break-up stage of multifragmentation.
 Hence it is possible to get back the value of the input symmetry energy coefficient using only the grand canonical model for calculating the yields of primary fragments at  the break-up stage.
The results of canonical and grand canonical ensembles differ in general for finite nuclei\cite{Mallik4} and hence in this case also the results from the two ensembles are different.

We have also extracted the value of $C_{sym}/T$ from the yields of projectile fragmentation reactions using our model for projectile fragmentation. The results from the primary fragments
are close to each other for all the four formulas used.
 In this model, where the canonical ensemble is used to calculate the yields of the hot primary fragments, the values of $C_{sym}/T$ obtained from the secondary fragments
 after evaporation are close to those obtained from the experimental yields but they differ from those  obtained from the primary fragments and from the input value of $C_{sym}$ used in the model.
  The message  we wish to convey
is that to deduce the value of the symmetry energy coefficient using the prescriptions described earlier [Eqs. (5) to (8)], it is advisable to use the grand
canonical model to obtain the yield of the fragments and one should use the yields at the break-up stage (primary fragments) of the reaction. The experimental yields (which are from the "cold" fragments)
should not be used to deduce the value of the symmetry energy coefficient since the formulas used for the deduction are all valid at the equilibrium stage or the breakup stage
of the reaction and secondary decay disturbs the equilibrium scenario of the breakup stage. Attempts to deduce the value of $C_{sym}$ from the secondary fragments or from the experimental
yields might lead to values that are very different from the actual value.\\

{\bf {\it Acknowledgement:-}}
The authors thank M. Mocko, Kelic-Heil Aleksandra and Karl-Heinz Schmidt for access to experimental data.

\end{document}